\documentclass{elsart}
\begin{document}
\begin{frontmatter}

\title{Static potential from spontaneous breaking of scale symmetry}
\author{Patricio Gaete\thanksref{cile}} 
\thanks[cile]{e-mail address: patricio.gaete@usm.cl}
\address{Departamento de F\'{\i}sica, Universidad T\'ecnica
Federico Santa Mar\'{\i}a, Valpara\'{\i}so, Chile}
\author{Eduardo Guendelman\thanksref{cile2}},
\thanks[cile2]{e-mail address: guendel@bgumail.bgu.ac.il }
\address{ Physics Department, Ben Gurion University, Beer
Sheva 84105, Israel}
\author{Euro Spallucci\thanksref{infn}}
\thanks[infn]{e-mail address: spallucci@ts.infn.it }
\address{Dipartimento di Fisica Teorica, Universit\`a di Trieste and INFN,
Sezione di Trieste, Italy}

\begin{abstract}
We determine the static potential for a heavy quark-antiquark pair
from the spontaneous symmetry breaking of scale invariance in a
non-Abelian gauge theory.  Our calculation is done  within the
framework of the gauge-invariant, path-dependent, variables
formalism. The result satisfies the 't Hooft  basic criterion for
achieving confinement.
\end{abstract}

\end{frontmatter}

\section{Introduction}

The binding energy of an infinitely heavy quark-antiquark pair
represents a key tool which is expected to play an important role in
the understanding of non-Abelian theories and especially of quark
confinement. In fact, the distinction between the apparently related
phenomena of screening and confinement have been of considerable
importance in order to gain further insight and guidance into this
problem.

As is well known, the problem of confinement in gauge theories has
been studied from different viewpoints, like lattice gauge theory
techniques \cite{Wilson} and non-perturbative solutions of
Schwinger-Dyson's equations \cite{Zachariesen,depple}.  We further
note that recently an interesting approach to this problem has been
proposed \cite{'t Hooft}, which includes a linear term in the
dielectric field that appears in the energy density. Mention should
be made, at this point, to the  "Cornell potential"  \cite{Eichten}
which simulates the features of $QCD$, that is,
\begin{equation}
V\left(\, r\,\right) =  - \frac{{\kappa}}{r} + \frac{r}{{a^2 }}\ , 
\label{Cornell}
\end{equation}
here $a$ is a constant with the dimensions of length. According to
't Hooft, confinement is associated to the appearance of a linear
term in the dielectric field ${\bf D}$ (that dominates for low
$|{\bf D}|$) in the energy density:
\begin{equation}
U({\bf D}) = \rho _{str}\, |{\bf D}|\ , \label{'t Hooft}
\end{equation}
the proportionality constant being the coefficient of the linear
potential, that is, $\rho _{str}  \approx \frac{1}{{a^2 }}$. It is
worthwhile remarking at this point that gauge theories with no scale
have a symmetry which is associated to this, scale invariance. Thus
it follows that the confinement phenomena breaks the scale
invariance as the Cornell potential (\ref{Cornell}) explicitly shows
by introducing the scale $a$.

With these ideas in mind, one of us (E.I.G.)
has studied the connection between scale symmetry breaking and
confinement in \cite{GaeteGuen2}. In effect, we have observed the appearance 
of the Cornell potential (\ref{Cornell}) as well as the 't Hooft relation
(\ref{'t Hooft}) after spontaneous breaking of scale invariance in
an Abelian model. As was explained in \cite{GaeteGuen}, the scale
invariant model under consideration introduces, in addition to the
standard gauge fields also maximal rank gauge field strengths of
four indices in four dimensions, $F_{\mu \nu \alpha \beta }  =
\partial _{\left[ \mu  \right.} A_{\left. {\nu \alpha \beta }
\right]}$ where $ A_{\nu\alpha\beta} $ is a three index potential.
The integration of the equations of motion of the $
A_{\nu\alpha\beta}$ field introduces a constant of integration $M$
which breaks the scale invariance \cite{cc}. More specifically, the linear
term in the Cornell potential arises from the constant of
integration $M$. Obviously, when $M=0$ the equations of motion
reduce to those of the standard gauge field theory.

Motivated by this, it is natural to ask wether a similar thing
happens in the case of a non-Abelian model with a spontaneously
broken symmetry. In this work we address this question and, as we
shall see, the confining potential between quark-antiquark emerges
naturally. Here we would mention that confinement arises as an
Abelian effect. In general, this picture agrees qualitatively with
that of Luscher \cite{Luscher}. More recently it has been related to
relativistic membrane dynamics in \cite{Gabadadze}, and implemented
through the \textit{abelian projection} method in \cite{Kondo}. We
further observe that some peculiar quantum aspects of the effective
long range dynamics of QCD, and certain intriguing analogies with
the Schwinger model, have been considered in \cite{Aurilia}. Once
again, our procedure fulfills completely the requirements by 't
Hooft for perturbative confinement. Our calculation is based on  the
gauge-invariant but path-dependent variables formalism \cite{Gaete1}
. According to this formalism, the interaction potential between two
static charges is obtained once a judicious identification of the
physical degrees of freedom is made.

\section{Scale symmetry breaking}

In this section, we discuss the connection between the scale
symmetry breaking and confinement, introduced in Ref.\cite{GaeteGuen}, in the
of path integral formulation. 
For this purpose we restrict our attention to the action
\begin{equation}
S_{YM} = \int {d^4 } x\left( - \frac{1}{4}F_{\mu \nu }^{a} F^{a\mu \nu}
 \right)\ ,
\end{equation}
where $ F_{\mu \nu }^a  = \partial _\mu  A_\nu ^a - \partial _\nu A_\mu ^a  
+ gf^{abc} A_\mu ^b A_\nu ^c$. We mention in passing that this theory is 
invariant under the scale symmetry
\begin{equation}
A^{a}_\mu  \left( x \right) \mapsto A_\mu ^ {a\prime}  \left( x
\right) = \lambda^{-1}\, A^{a}_\mu  \left( \lambda\, x \,\right)\ ,
\end{equation}
where $\lambda$ is a constant.

Let us introduce  the following \textit{parent action} functional \cite{dual}
encoding information about the different phases of the model
\begin{eqnarray}
{\mathcal{S}_P}\left[\, \omega\ ,\phi\ ,  A^{a}_\mu\ , A_{\mu\nu\rho}\,\right]
 =&& \int {d^4 } x\left[\,  - \frac{1}{4}\omega ^2  
+\frac{1}{2}\omega \sqrt { - F_{\mu \nu }^{a}\, F^{a\mu \nu } } 
\right.\nonumber\\
+&&  \phi(x)\,\left.\left(\, \frac{1}{4!}\epsilon^{\lambda\mu\nu\rho}
\partial_{[\,\lambda}\, A_{\mu\nu\rho \,]}
-\omega\,\right)
\,\right]\ ,
\end{eqnarray}
where, $\omega$ and $\phi$ are auxiliary scalar fields, and $A_{\mu\nu\rho}$ 
is an antisymmetric, rank-three, gauge form. Their scaling transformations are
\begin{eqnarray}
&&\omega  \mapsto \lambda ^{-2} \omega \left(\, \lambda\, x\right)\label{cr25}\\
&& \phi \mapsto \lambda ^{-2} \phi\left( {\lambda x}\right)\label{cr25b}\\
&&A_{\mu\nu\rho}\mapsto \lambda ^{-1} A_{\mu\nu\rho}\left( {\lambda x}\right)
\label{cr25c}
\end{eqnarray}

Extended gauge potential have an ubiquitous role in theoretical high energy 
physics. They
act as gauge partner of relativistic extended objects of various dimensions 
\cite{ant},  and play a fundamental role in cosmology as well \cite{cosmo}. 
It can be worth to recall that
in Yang-Mills theory rank-three gauge potential enters through the topological 
density term  $F^\ast{}_{\mu \nu }^{a}\, F^{a\mu \nu }$ and couples to the 
membrane boundary of hadronic bubble \cite{castro}. As an abelian, colorless 
object, $A_{\mu\nu\rho}$ is expected 
to describe the long-range ``tail'' of $QCD$ \cite{Luscher},\cite{Gabadadze}.\\  
We notice that the Yang-Mills field enter the action through a Born-Infeld-like
 term in the strong field limit, suggesting a possible connection with 
 strings/brane  formulation of gauge theory  \cite{bi},\cite{Amer}.\\
Dynamics of the whole system is encoded into the partition functional

\begin{equation}
Z\equiv \mathcal{N}\int \left[\, \mathcal{DA}^a_\mu\,\right]\,
\left[\,\mathcal{D\omega}\,\right]\,\left[\,\mathcal{D\phi}\,\right]\,
\left[\,\mathcal{DA_{\lambda\mu\nu\rho}}\,\right]\, \exp\left(\,
-{\mathcal{S}_P}\left[\, \omega\ ,\phi\ ,  A^{a}_\mu\ , A_{\mu\nu\rho}\,\right]
\,\right)
\end{equation}
where $\mathcal{N} $ is a suitable normalzation constant.\\
Suppose we start by integrating out the three-form gauge potential. We get
the following ``constrained'' partition functional

\begin{eqnarray}
Z\equiv \mathcal{N}\int\, \dots\, &&\left[\,\mathcal{D\phi}\,\right]\,
\delta\left[\, \epsilon^{\lambda\mu\nu\rho}
\partial_\lambda \phi\,\right]\,\times\nonumber\\
&& \exp\left(\,-\int d^4x\,\left[\,     
 - \frac{1}{4}\omega ^2  +
\frac{1}{2}\omega \sqrt { - F_{\mu \nu }^{a}\, F^{a\mu \nu } } 
-\phi(x)\,\omega\,\right)\,\right]
 \end{eqnarray}

The Dirac-delta restricts the functional integration over $\phi$ to 
constant field configurations only:

\begin{equation}
\int\,\left[\,\mathcal{D\phi}\,\right]\,
\delta\left[\, \epsilon^{\lambda\mu\nu\rho}
\partial_\lambda \phi\,\right]\left(\,\dots\,\right)=
\int\left[\,\mathcal{D\phi}\,\right]\,\delta\left[\, 
 \phi -M\,\right]\,\left(\,\dots\,\right)
\end{equation}

Thus, we find

\begin{eqnarray}
Z_M =\mathcal{N}\int &&\left[\,\mathcal{DA}^a_\mu\,\right]\, 
\left[\,\mathcal{D\omega}\,\right]\,\times\nonumber\\
&&\exp\left(\,-\int d^4x\,\left[\, 
     - \frac{1}{4}\omega ^2  +
\frac{1}{2}\,\omega \left(\,\sqrt { - F_{\mu \nu }^{a}\, F^{a\mu \nu } } 
-M\,\right)\,\right]\,\right)
\end{eqnarray}
the integration constant  has $(mass)^2$ dimension in natural units. 
The appearance of a dimensional constant signals the breaking of scale 
invariance. 
The $M$ constant acts as an \textit{order parameter} for
the different phases of the system. If $M=0$ we recover the familiar
Yang-Mills theory after integration over the $\omega$ field:

\begin{eqnarray}
Z_{ M=0} &&=\mathcal{N}\int \left[\,\mathcal{DA}^a_\mu\,\right]
\,\left[\,\mathcal{D\omega}\,\right]\,
\exp\left(\,-\int d^4x\,\left[\, 
     - \frac{1}{4}\omega ^2  +
\frac{1}{2}\,\omega \left(\,\sqrt { - F_{\mu \nu }^{a}\, F^{a\mu \nu } } 
\,\right)\,\right]\,\right)\nonumber\\
&&=\mathcal{N}\int\,\left[\, \mathcal{DA}^a_\mu\,\right]\,
\exp\left(\,-\int d^4x\,\left[\, 
    -\frac{1}{4} F_{\mu \nu }^{a}\, F^{a\mu \nu }  \,\right]\,\right)
    \equiv
 Z_{YM}
\end{eqnarray}

On the other hand, if $M\ne 0$ we get an additional Born-Infeld contribution
to the Yang-Mills action, coming from the breaking of scale invariance
\cite{GaeteGuen2}

\begin{equation}
Z_M = 
\int d^4x\,\left[\,  - \frac{1}{4} F_{\mu \nu }^{a}\, F^{a\mu \nu }  
+\frac{M}{2}\, \sqrt { - F_{\mu \nu }^{a}\, F^{a\mu \nu } } 
\,\right],
\end{equation}

where field-independent constant has been absorbed into the
normalization factor $\mathcal{N}$. In this case the
action contains both a Yang-Mills and a Born-Infeld term.
It is important to realize that the integration constant $M$
spontaneously breaks the scale invariance, since both $\omega$ and
$\sqrt { - F^{a\mu \nu } F^{a}_{\mu \nu } }$ transform as in
Eq.(\ref{cr25}) but $M$ does not transform. We also call attention
to the fact that $M$ has the same dimensions as the field strength
$F^{a}_{\mu\nu}$, that is, dimensions of $\left( {length} \right)^{
- 2}$. It now follows that the variation of the $A^{a}_\mu$ field
produces the following equation
\begin{equation}
 \nabla _ \mu
\left[ {\left( {\sqrt { - F_{\alpha \beta }^a F^{a\alpha \beta } } +
M} \right)\frac{{F^{a\mu \nu } }}{{\sqrt { - F_{\alpha \beta }^b
F^{b\alpha \beta } } }}} \right] = 0\ . \label{cr50}
\end{equation}
Once again, we observe as an appealing feature of this expression,
that the introduction of the unusual $M$ term leads to the
generation of confinement. 
The above equation admits a ``trivial vacuum'' solution for $M=0$, which is

\begin{eqnarray}
&& M=0\\
&& F^a_{\mu\nu}=0
\label{trivial}
\end{eqnarray}

while, for $M\ne 0$ one finds that the classical field strength is subject to
the condition

\begin{equation}
\sqrt{- F_{\alpha\beta }^b \, F^{b\alpha \beta } }	=  -M \ . \label{cr51}
\label{fquadro}
\end{equation}

This kind of constrained  field configurations has been originally
introduced in \cite{nambu} in order to proivide a gauge type description
of string dynamics. Later in \cite{bi},\cite{Amer} an in-depth investigation
of gauge-type formulation of string theory has been given. It may be worth
to recall that the \textit{effective string tension} is given by \cite{bi}:

\begin{equation}
\frac{1}{2\pi\alpha^\prime}= \frac{\vert\, M\, \vert}{\sqrt 2}
\end{equation}

Confinement is obvious
then, since in the presence of two external sources, by symmetry
arguments, one can see that such a constant amplitude
chromoelectric field must be in the direction of the line joining
the two charges. The potential that gives rise to this kind of
field configuration is of course a linear potential.\\
The two solutions (\ref{trivial}) and (\ref{fquadro}) clearly
show in which sense scale invariance is ``spontaneously broken'' and
how this effect leads to linear confinement.
Notice that this equation  gives confinement only if $M<0$, otherwise the 
chromoelectric field will be antiparallel to itself, giving that the 
confinement piece is zero.
Accordingly, the term inside the square
brackets corresponds to the ''dielectric field $D^{a\mu \nu }$''.
Hence we see that the equations of motion would be obtained from an
action of the form
\begin{equation}
S = k\int d^4 x\sqrt { - F_{\mu \nu }^{a} F^{a\mu \nu } } \ ,
\label{cr55}
\end{equation}
where $k$ is a constant. Such model leads to confinement and to
string solutions \cite{Aurilia},\cite{Amer}.\\
In order to compute the interaction energy we need to write
the Born-Infeld term in a more tractable form. To achieve this goal
we need to introduce an auxiliary field, say $e(x)$. Thus, we get
the on-shell equivalent form
\begin{equation}
S\left[\, A_\mu^a\ ,e(x)\,\right]=
 \int d^4 x\left[\, -\frac{M}{4} e(x) F_{\mu \nu }^{a} F^{a\mu \nu }
+ \frac{M}{4} \frac{1}{e(x)}\,\right]\ ,
\end{equation}

\begin{equation}
S\left[\, A_\mu^a\ ,e(x)\,\right]=S_{BI}\left[\, A_\mu^a\,\right]
\longleftrightarrow e(x)= \frac{1}{\sqrt { - F_{\mu \nu }^{a} F^{a\mu \nu } }}
\ .
\end{equation}

By adding the Yang-Mills term we get for the total action
\begin{equation}
{\mathcal{S}_P}\left[\, A_\mu^a\ ,e(x)\,\right]= \int d^4 x\left[\, -\frac{1}{4}
\left(\, 1 + Me(x)\,\right) F_{\mu \nu }^{a} F^{a\mu \nu }
+ \frac{M}{4} \frac{1}{e(x)}\,\right]\ . \label{cr60}
\end{equation}

The action (\ref{cr60}) explicitly shows that we obtained a
Yang-Mills theory with an effective, \textit{point-dependent},
 dielectric constant
 
 \begin{equation}
 \varepsilon_{YM}(x) \equiv  1 + M e(x)\ .
 \label{diel}
 \end{equation}

Equation (\ref{diel}) is a consequence of the non-trivial properties
of the Yang-Mills vacuum parametrized through the function $e(x)$.\\
To conclude this section we notice that the dimension of $M$ are
the same as the string tension:

 \begin{equation}
 \left[\, M\,\right]=\left[\, 1/\alpha^\prime\,\right]
\end{equation}

Thus, we expect that the appearance of such a constant is a signal
of a \textit{stringy} phase of the Yang-Mills field. In the next section
we are going to show this expectation is correct, as $M$ enters
in the linear part of the interaction energy between a pair of static
test charges.

\section{Interaction energy}

We now examine  the interaction energy between external probe
sources in the model (\ref{cr60}). This can be done by computing the
expectation value of the energy operator $H$ in the physical state
$\left| \Phi  \right\rangle$, which we will denote by $\left\langle
H \right\rangle _\Phi$. The model (\ref{cr60}) cannot be analytically solved 
in full generality. Thus, we impose spherical symmetry and reduce the problem 
to an effective $1+1$ dimensional model,
where the interaction potential can be exactly determined.\\
Our starting point is the Lagrangian 
density (\ref{cr60}) 
\begin{equation}
\mathcal{L} = 4\pi r^2 \left\{ { - \frac{1}{4}\left( {1 + Me} \right)
F_{\mu \nu }^a F^{a\mu \nu }  + \frac{M}{4}\frac{1}{e}} \right\} 
- A_0^a J^{a0}\ , \label{inter5}
\end{equation}
where $J^{a0} $ is the external current. In passing we note that $\mu,\nu=0,1$ 
and $ x^1  \equiv r \equiv x$. As we have noted before, by introducing the 
auxiliary field $\varepsilon_{YM}(x) \equiv\frac{1}{V}=1+Me$, 
expression  (\ref{inter5}) then becomes
\begin{equation}
\mathcal{L} = -4\pi r^2 \left\{  \frac{1}{4V}F_{\mu \nu }^a 
F^{a\mu \nu }  + \frac{{M^2 }}{4}\frac{V}{{V - 1}} \right\} - A_0^a J^{a0}\ . 
\label{inter5a}
\end{equation}

Once this is done, the canonical quantization of this theory from
the Hamiltonian analysis point of view is straightforward. The canonical
momenta read $\Pi ^{a\mu }  =  - 4\pi x^{2}\frac{1}{V}F^{a0\mu }$, and one 
immediately identifies the two primary constraints $\Pi^{a0}=0$ and 
$P \equiv \frac{{\partial \mathcal{L}}}{{\partial  \dot V }} = 0$. Standard 
techniques for constrained systems then lead to the following canonical 
Hamiltonian:
\begin{eqnarray}
H_C && = \int {dx} \left(  - \frac{V}{8\pi x^2 }\,\Pi _i^a \,
\Pi ^{ai}  + \Pi _i^a \,\partial ^i\, A^{a0} \right.\nonumber\\
  && - gf^{abc} \left. \Pi _i^a A^{b0} A^{ci}+ \frac{1}{4}F_{ij}^a F^{aij}
 + A_0^a J^{a0}  + 4\pi x^2 \frac{{M^2 }}{4}\frac{V}{V - 1}\, \right)\ . 
 \label{inter10}
\end{eqnarray}

Time conserving the primary constraints $\Pi^{a0}\approx0$ yields
the secondary constraints $\Gamma ^{a \left( 1 \right)} \left( x
\right) \equiv \partial _i \Pi ^{ai}  + gf^{abc} A^{bi} \Pi _i^c  -
J^{a0}  \approx 0$. The consistency condition for the $P$ constraint yields no 
further constraints and just determines the field $V$,
\begin{equation}
V = 1 + \frac{|M|}{{\sqrt 2 }}4\pi x^2 \frac{1}{{\sqrt {\Pi ^{ai} \Pi ^{ai} } }}
\ , \label{inter10a}
\end{equation}
which will be used to eliminate $V$. The extended Hamiltonian that generates 
translations in time then reads 
$H = H_C  + \int d x \left( {c_0^{a} (x)\Pi_0^{a} (x) + c_1^{a} (x)
\Gamma ^{a \left( 1 \right)} \left( x \right)} \right)$, where $c_0^{a}(x)$ 
and $c_1^{a}(x)$ are the Lagrange multipliers. Moreover, it follows from this 
Hamiltonian that $ \dot{A}_0^{a} \left( x \right) = 
\left[ {A_0^{a} \left( x \right),H} \right] =
c_0^{a} \left( x \right)$, which are arbitrary functions. Since
$\Pi^{0a} = 0 $, neither $ A^{0a}$ nor $\Pi^{0a}$ are of interest in
describing the system and may be discarded from the theory. The
Hamiltonian then reads
\begin{equation}
H = \int {dx} \left\{ {\frac{{\Pi ^{ai} \Pi ^{ai} }}{{8\pi x^2 }} 
+ \frac{|M|}{{\sqrt 2 }} + \frac{1}{4}F_{ij}^a F^{aij}  
+ c^a \left( {\partial _i \Pi ^{ai}  + gf^{abc} A^{bi} 
\Pi _i^c  - J^{a0} } \right)} \right\}, \label{inter15}
\end{equation}
where $c^a  \left( x \right) = c_1^{a} \left( x \right) - A_0^{a}
\left( x \right)$.

According to the usual procedure we introduce a supplementary
condition on the vector potential such that the full set of
constraints becomes second class. A useful choice is found to be
\cite{Gaete1}:
\begin{equation}
\Gamma^{a\left( 2 \right)} \left( x \right) = \int\limits_0^1
{d\lambda } \left( {x - \xi } \right)^i A_i^{\left( a \right)}
\left( {\xi  + \lambda \left( {x - \xi } \right)} \right) \approx
0, \label{inter20}
\end{equation}
where  $\lambda$ $(0\leq \lambda\leq1)$ is the parameter describing
the spacelike straight path $ x^i  = \xi ^i  + \lambda \left( {x -
\xi } \right)^i $, and $ \xi $ is a fixed point (reference point).

This supplementary condition is the non-Abelian generalization of the gauge 
condition discussed in \cite{GaeteDu}, which leads to the Poincar\'{e} gauge. 
 There is no essential loss of generality if we restrict our considerations to 
 $ \xi ^i=0 $. In this case, the only nontrivial
Dirac bracket is
\begin{eqnarray}
\left\{ {A_i^a \left( x \right),\Pi ^{bj} \left( y \right)}
\right\}^ *   &&= \delta ^{ab} \delta _i^j \delta ^{(3)} \left( {x -
y} \right)\nonumber\\ 
&&- \int\limits_0^1 {d\lambda } \left( {\delta ^{ab}
\frac{\partial }{{\partial x^i }}
 - gf^{abc} A_i^c \left( x
\right)} \right)x^j \delta ^{(3)} \left( {\lambda x - y} \right)\ . 
\label{inter25}
\end{eqnarray}
In passing we note the presence of the last term on the right-hand
side which depends on $g$.

Now we are in a position to be able to compute the interaction
energy between pointlike sources in the theory under consideration,
where a fermion is localized at the origin $ {\bf 0}$ and an
antifermion at $ {\bf y}$. As we have already indicated, we
will calculate the expectation value of the energy operator $ H$ in
the physical state $|\Phi\rangle$. From our above discussion we then 
get for the expectation value 
\begin{equation}
\left\langle\, H\, \right\rangle _\Phi   = \mathrm{Tr}\left\langle\, \Phi\,  
\right\vert\, \int dx\, \left(\, \frac{\Pi ^{ai} \Pi ^{ai} }{8\pi x^2 } 
+ \frac{|M|}{{\sqrt 2 }}\sqrt{\, \Pi ^{ai} \Pi ^{ai} } \, 
+ \frac{1}{4}F_{ij}^a F^{aij} \right)\left\vert \Phi  
\right\rangle \ . \label{inter30}
\end{equation}
It is easy to see that the first term inside the bracket comes from the usual 
Yang-Mills theory while the second one is a correction which comes from the 
square root modification.

Now we recall that the physical state can be written as \cite {Gaete1},
\begin{equation}
\left| \Phi  \right\rangle  =
\overline \psi  \left( {\bf y} \right)U\left( {{\bf y},{\bf 0}} \right)
\psi \left( {\bf 0} \right)\left| 0 \right\rangle\ , \label{inter35}
\end{equation}
where
\begin{equation}
U\left( {{\bf y},{\bf 0}} \right) \equiv P\exp \left( {ig\int_{\bf 0}^{\bf y}
{dz^i A_i^a \left( z \right)T^a } } \right)\ . \label{inter40}
\end{equation}
As before, the line integral is along a spacelike path on a fixed
time slice, $P$ is the path-ordering prescription and $\left|
0\right\rangle$ is the physical vacuum state. As in \cite{Gaete1},
we again restrict our attention to the weak coupling limit.

From this and the foregoing Hamiltonian discussion, we then get
\begin{equation}
\left\langle H \right\rangle _\Phi   = \left\langle H \right\rangle
_0  + V_1  + V_2, \label{inter45}
\end{equation}
where $\left\langle H \right\rangle _0  = \left\langle 0
\right|H\left| 0 \right\rangle$, and the $V_1$ and $V_2$ terms are
given by:
\begin{equation}
V_1  = \mathrm{Tr}\left\langle \Phi  \right|\int dx\,\left(\,
\frac{{ \Pi ^{ai}  \Pi ^{ai} }}{8\pi\, x^2 }
\,\right)\left| \Phi \right\rangle\ , \label{inter50}
\end{equation}
and
\begin{equation}
V_2  = \frac{|M|}{{\sqrt 2 }}\,\mathrm{Tr}\left\langle \Phi  \right|\int {dx} 
\sqrt{\,\Pi ^{ai} \Pi ^{ai} } \left| \Phi  \right\rangle\ . \label{inter55}
\end{equation}

For more technical details about the derivation of (\ref{inter45}) we
refer to the the following papers \cite{consist}.\\
As we have noted before, the $V_{1}$ term is similar to the energy
for the Yang-Mills theory. Nevertheless, in order to put our
discussion into context it is useful to summarize the relevant
aspects of the analysis described previously \cite{Gaete1}. From
(\ref{inter50}) we then get an Abelian part (proportional to
$C_{F}$) and a non-Abelian part (proportional to the combination
$C_{F}C_{A}$). As we have noted before, the Abelian part takes the
form
\begin{equation}
V_1^{\left( {g^2 } \right)}  =   \frac{g^2}{2} \mathrm{Tr}\left(\, {T^a T^a }
\,\right)\, \int_0^y {dz^1 } \int_0^y d {z^ \prime}{}^1 \,\frac{{\delta
\left( {z^1 - z^{\prime ^1 } } \right)}}{{4\pi \left( {z^1 }
\right)^2 }}\ , \label{inter60}
\end{equation}
Writing the group factor $\mathrm{Tr}(T^{a}T^{a})=C_{F}$, the expression
(\ref{inter60}) is given by
\begin{equation}
V_1^{(g^2) } \left( L \right) =  - \frac{1}{{8\pi }}g^2 C_F
\frac{1}{L}\ , \label{inter65}
\end{equation}
where $\left|y \right| \equiv L$. Next, the non-Abelian part may be
written as
\begin{equation}
V_1^{\left( {g^4 } \right)}  = \mathrm{Tr}\int {\frac{{dx}}{{8\pi x^2 }}}
\left\langle 0 \right|\left( {I^1 } \right)^2 \left| 0
\right\rangle, \label{inter70}
\end{equation}
where
\begin{equation}
I^1  = g^2 f^{abc} T^b \int_0^y {dz^1 } \int_0^1 {d\lambda } A_1
\left( {z^1 } \right)z^1 \delta \left( {x^1  - \lambda z^1 }
\right)\ . \label{inter75}
\end{equation}
Expression (\ref{inter70}) reduces to
\begin{equation}
V_1^{(g^4) } \left(\, L\, \right) = \frac{1}{4}g^4 C_A C_F \left( { -
\frac{1}{L}} \right)\int_{0}^{y} {dz^1 } \int_{\bf 0}^{\bf y} {dz^{
\prime 1} } D_{11} \left( {z^1,z^ {\prime 1} } \right)
\label{inter80}
\end{equation}
Here $D_{11}(z^1,z^{\prime 1})$ stands for the propagator, which is
diagonal in color and taken in an arbitrary gauge. Following our
earlier discussion, we choose the Feynman gauge. As a consequence,
expression (\ref{inter80}) then becomes
\begin{equation}
V_1^{(g^4) } \left(\, L\, \right) =  - g^4 \frac{1}{{8\pi ^2 }}C_A C_F
\frac{1}{L}\log \left( {\mu L} \right)\ , \label{inter85}
\end{equation}
where $\mu$ is a cutoff. Then, the $V_{1}$ term takes the form
\begin{equation}
V_1  =  - g^2 C_F \frac{1}{{8\pi L}} \left( {1 + \frac{{g^2 }}{\pi
}C_A \log \left( {\mu L} \right)} \right)\ . \label{inter90}
\end{equation}

The task is now to evaluate the $V_{2}$ term, which is given by
Eq.(\ref{inter55}). Since the field $\Pi^{ai}$ is basically the electric 
field $F^{a0i}$, we restrict ourselves to constant electric fields 
(in color space)  in order to handle the square root in expression 
(\ref{inter55}). In such a case, we can write $\Pi^{ai}=v^{a}\Pi^{ai}$, 
where $v^{a}$ is a constant vector in color space.  In this way, color 
invariance is explicitly broken. The presence of a constant chromomagnetic
field breaks Lorentz invariance as well.
However, both symmetries are 
recovered at large distance by averaging over a large number of randomly 
oriented vacuum domains. In any case, the confining features of the model are 
not affected by the averaging procedure. In fact, this same aspect has been 
discussed recently in the context of $(2+1)$ and $(3+1)$ - dimensional 
reformulated $SU(2)$ Yang-Mills theory \cite{GGS}.

In this case, the only nontrivial contribution is the Abelian one, that is,
\begin{equation}
V_2  = \frac{|M|}{{\sqrt 2 }}g\,\mathrm{Tr}\left(\, {v^{1a} e^1 T^a }\, \right)\,
L\ .
\label{inter95}
\end{equation}
where $e^1$ is a unit vector along the $z^1$.

By putting together equations (\ref{inter90}) and (\ref{inter95}),
we obtain for the total interquark potential
\begin{equation}
V =  - g^2 \, C_F \frac{1}{8\pi L}\left( {1 + \frac{{g^2 }}{\pi }\, C_A
\log \left( {\mu L} \right)} \right) + \frac{{|M|g}}{{\sqrt 2
}}\mathrm{Tr}\left( {v^{1a} e^1 T^a } \right)\, L\ , \label{inter100}
\end{equation}
which has the Cornell form. Notice that confinement is obtained at a
finite value of the strong coupling just as claimed by 't Hooft
\cite{'t Hooft}.

\section{Conclusions}

Let us summarize our work. We have shown that the non-Abelian
generalization of the model considered in Ref. \cite{GaeteGuen}
leads to the Cornell confining potential. More interestingly, it was
found that confinement is basically an Abelian effect. In general,
this picture agrees qualitatively with that of Luscher
\cite{Luscher}.  Once again, the gauge-invariant formalism has been
very economical in order to obtain the interaction energy, this time
showing a confining effect in the context of a non-Abelian effective
model. We also draw attention to the fact that the model satisfies
indeed the 't Hooft basic criterion for achieving confinement, even
with finite coupling constant. In such a case, the necessary term
for the dependence of the energy density for low dielectric field
(linear in $|{\bf D}|$) discussed by 't Hooft is here obtained as a
result of spontaneous breaking of scale invariance which introduces
the constant $M$.

\section{ACKNOWLEDGMENTS}

One of us (P.G.) wants to thank the Physics Department of the
Universit\`a di Trieste for hospitality. Work supported in part by Fondecyt 
(Chile) grant 1050546 (P.G.).

\end{document}